\documentclass{article}

\usepackage{amssymb}
\usepackage{amsfonts}
\usepackage{amsmath}
\usepackage{subfigure}
\usepackage{graphicx,lscape}
\usepackage{picinpar}

\input{tcilatex}

\begin{document}

\title{Quantum Information, Thermofield Dynamics and Thermalized Bosonic Oscillator}
\author{M. A. S. Trindade$^{1,3}$\footnote{matrindade@uneb.br}, L. M. Silva Filho$^{1}$\footnote{lourivalmsfilho@uol.com.br}, L. C.
Santos $^{2,3}$\footnote{leandro.c@ufrb.edu.br}, \\
M. Gra\c{c}as R. Martins $^{3}$\footnote{mgraca@ufba.br}, J. D. M. Vianna$^{3,4}$\footnote{jdavid@fis.unb.br}\\
\ {\small 1. Departamento de Ci\^{e}ncias Exatas e da Terra II, UNEB, Alagoinhas-Ba, Brasil} \\
\ {\small 2. Centro de Ci\^encias Exatas e Tecnologia, UFRB, Cruz das Almas-Ba, Brasil}\\
\ {\small 3. Instituto de F\'{\i}sica, UFBA, Salvador-Ba, Brasil}\\
\ {\small 4. Instituto de F\'{\i}sica, UnB, Bras\'{\i}lia-DF, Brasil.}}

\maketitle

\begin{abstract}
We show through Thermofield Dynamics approach that the action of the thermalized quantum logic gate  on the thermalized state is equivalent to thermalization of the state that arise from the application of the non-thermalized quantum logic gate. In particular, we study the effect of temperature on a mixed state associated to a system capable of implementing a \textit{CNOT} quantum logic gate. According to a proposal in the literature, a way of implementing such a logic gate is by using a representation of the qubit states as elements of the Fock space of a bosonic system. We consider such a proposal and  use the Thermofield Dynamics to determine the thermalized  qubit  states. The temperature acts as a quantum noise on pure states, making them a statistical mixture. In this context we analyze the fidelity as a function of  the temperature and using the Mandel parameter, we determine temperature ranges  for which the statistics of the system  becomes subpoissonian, poissonian and superpoissonian. Finally, we calculate the Wigner function, allowing an analysis of the thermal state in phase space, and we obtain that  the increase of temperature decreases non-classical properties of the system. The temperature range where one has a subpoissonian statistics and  high fidelity is determined.

\end{abstract}

\section{Introduction}

Since the last decades a large number of theoretical proposals \cite{Lloyd, Chuang, Cirac, DDiVicenzo, DLoss, Santana, Ekij, Sharma} for physical implementation of quantum states (the qubits) has been performed. These proposals specially take as their foundation the possibility of physical implementation of quantum logic gates. Among them one of the most used is the $CNOT$ (controlled-NOT) gate since it is known that any unitary computing operation can be composed from the spin and the $CNOT$ gates \cite{DiVicenzo}, a result commonly known in the literature as universal property. The decoherence, on the other hand, has been the major obstacle \cite{NilesenM, Lidar},  in the experimental implementation of quantum computing. No classical or quantum system is perfectly isolated so that a noise source can affect the state of qubit corrupting the desired evolution of the system. So, a system initially prepared in a pure state in such a way that it is possible to perform a given logic gate, can evolve to other state than the original, making operation unsuccessful. The depolarization channel \cite{Klimov}, for example, transforms the pure state $\left\vert \psi \right\rangle \left\langle \psi \right\vert $ in the mixed state $\varepsilon (\rho )=\frac{p}{2}I+(1-p)\left\vert \psi \right\rangle \left\langle \psi \right\vert $; other well-studied channel, the GAD (generalized-amplitude-damping) \cite{Davidovich, TingYu}, usually associated with systems in a thermal bath with arbitrary temperature, has influence on quantum entanglement. In reference \cite{Masashi} Masashi Ban analyzed the dephasing of interacting two qubits under the influence of independent thermal reservoirs, and using the measure of concurrence examined the decay of entanglement for the two qubits. Ali \emph{et al} \cite{Mazhar} presented a criterion that characterizes the conditions for sudden death of entanglement (ESD) of X-states of two qubits interacting with statistically uncorrelated thermal reservoirs; they proved that if at least one of the reservoirs is at finite temperature, all X-states exhibit ESD. Banerjee \cite{Bonerjee} conducted a study of the noise effect on the entanglement generated between two spatially separated qubits by considering their interaction with a squeezed-thermal bath; the entanglement birth and death were revealed through an analysis of concurrence in localized decoherence model. The temperature in dynamics of atom + cavity system has been explored using Thermofield Dynamics in reference \cite{Matra}, showing that the dynamics of the system is sensitive to small changes in the temperature

These observations suggest that it is interesting  to investigate the influence of temperature on the nonclassical properties of systems with interest for quantum computing. In fact, quantum features can be modified due to temperature, decreasing the level of non-classicality of the systems, for instance. This can be inferred directly from an analysis of the Mandel parameter, observing as the subpoissonian statistical depends on the temperature and, in the case of the Wigner function, by observing its negative values. In the present communication we analyze these questions by using the Thermofield Dynamics in order to introduce the temperature in the study of qubit states. Specifically, we will investigate the effect of temperature on a mixed state associated to a bosonic system capable of implementing the $CNOT$ quantum logic gate. In this context we analyze the fidelity and non-classical properties of the system using the Mandel parameter and the Wigner function.

The material of this paper is organized as follows. In Sec.2 we present some basic elements of the Thermofield Dynamics. In Sec.3 the representation of the qubit states as elements of the Fock space of a bosonic system is given, the correspondent thermalized states are obtained and the density operator determined. In Sec.4 we analyze the fidelity which allows to us a comparison between the original pure state $|\psi\rangle $ and the thermalized state $|\psi (\beta )\rangle$. In Sec.5 we present the calculations of the Mandel parameter and Sec.6 contains the Wigner function and its behavior with temperature. In Sec.7 we make our conclusions.

\section{ Elements of Thermofield Dynamics}

The effect of temperature on quantum systems can be studied via a formalism developed by Umezawa \cite{Takahashi,Umezawa, David, Ademir, David1, liethermo, Ying, Thiago}, called Thermofield Dynamics. This formalism has as its starting point the equality
\begin{equation}
\left\langle 0\left( \beta \right) \right\vert A\left\vert 0\left( \beta \right) \right\rangle = Tr(1/Z\exp(-\beta H)A),  \label{eq1}
\end{equation}
where the first term is the usual average for a pure state in quantum mechanics but considering the temperature ($\beta=\frac{1}{\kappa_B T}$, where T is temperature and $\kappa_B$ is the Boltzmann constant) and the second one corresponds to the average for statistical mixture; $A$ is an observable, $H$ is the Hamiltonian operator and $Z$ is the partition function. The establishment of this equality implies the doubling of the Hilbert space $ \mathcal{H}$ of the original system in order to treat the mixed state $\rho =1/Z\exp (-\beta H)$ as a pure state in the doubled Hilbert space $\mathcal{H}\otimes \tilde{\mathcal{H}}$. In eq.(\ref{eq1}) the state $|0(\beta )>$ is
\begin{equation}
\left\vert 0\left( \beta \right) \right\rangle =U(\beta )\left\vert 0,\overset{\sim }{0}\right\rangle ,
\end{equation}
where $U(\beta )$ is the Bogoliubov transformation

\begin{equation}
U(\beta )=\exp [-i\theta (\beta )(\tilde{a}a-\tilde{a}^{\dagger }a^{\dagger})],
\end{equation}\label{eq4}
that takes the non-thermal vacuum $|0,\tilde{0}>$ in the true vacuum state $|0(\beta )>$. In \ eq.(3) $a$, $a^{\dagger }$ ($\tilde{a}$, $\tilde{a}^{\dagger }$) are bosonic annihilation and creation operators in $\mathcal{H}$ (in $\mathcal{\tilde{H}}$), respectively, and $\theta (\beta )$ is an hyperbolic angle. Considering a bosonic oscillator the density matrix associated with the thermal vacuum $|0(\beta )>$ is given by \cite{Ademir}

\begin{eqnarray}
\rho _{\left\vert 0\left( \beta \right) \right\rangle } &\equiv &\rho_{\beta }=\frac{1}{1+\overset{-}{n}(\beta )}\underset{n=0}{\overset{\infty }{\sum }}\left( \frac{\overset{-}{n}(\beta )}{1+\overset{-}{n}(\beta )}\right)
^{n}\left\vert n\right\rangle \left\langle n\right\vert   \notag \\
&=&k\underset{n=0}{\overset{\infty }{\sum }}k_{1}^{n}\left\vert
n\right\rangle \left\langle n\right\vert ,  \label{rdezerobeta}
\end{eqnarray}
with $\overset{-}{n}(\beta )=\frac{1}{e^{\beta \omega }-1},k=\frac{1}{1+\overset{-}{n}(\beta )}$ and $k_{1}=\left( \frac{\overset{-}{n}(\beta )}{1+\overset{-}{n}(\beta )}\right)$. We obtain easily that for a general thermalized state  $|\psi (\beta )\rangle $ in Hilbert space

\begin{equation}
\left\vert \Psi \left( \beta \right) \right\rangle =f(a,a^{\dagger };\beta
)\left\vert 0\left( \beta \right) \right\rangle,
\end{equation}
and the associated density operator is

\begin{equation}
\rho _{\left\vert \Psi \left( \beta \right) \right\rangle }=f(a,a^{\dagger
};\beta )\rho _{\beta }f^{\dagger }(a,a^{\dagger };\beta ).
\end{equation}

In this article, based on the formalism of Thermofield Dynamics, we will investigate the effect of temperature on a mixed state associated to a bosonic system, capable of implementing the \textit{CNOT} quantum logic gate. In this context we analyze the fidelity and non-classical properties of the system using the Mandel parameter and the Wigner function.

\section{Representation of qubits and thermalized states}
Consider first a quantum logic gate $U_{L}$ that operate in the state $\vert {\psi^{'}}\rangle$ result in the state $\vert {\psi}\rangle$, i.e.,
\begin{equation}
U_{L}\vert {\psi^{'}}\rangle=\vert {\psi}\rangle
\end{equation}
We can show that the action of the thermalized logic gate $U(\beta)$ on the thermalized state $\vert {\psi^{'}(\beta)}\rangle$ is equivalent to thermalization of the state $\vert {\psi,\overset{\sim }{0}}\rangle$ that arise from the application of the non-thermalized logic gate. In fact,
\begin{eqnarray}
U_{L}(\beta)\vert {\psi^{'}(\beta)}\rangle&=&U(\beta)U_{L}U^{\dag}(\beta)\vert {\psi^{'}(\beta)}\rangle \notag\\
                                          &=&U(\beta)U_{L}U^{\dag}(\beta)U(\beta)\vert {\psi^{'},\overset{\sim }{0}}\rangle \notag\\
                                          &=&U(\beta)U_{L}\vert {\psi^{'},\overset{\sim }{0}}\rangle \notag\\
                                          &=&U(\beta)\vert {\psi,\overset{\sim }{0}}\rangle \label{nova1}
\end{eqnarray}
Following the proposal of Lloyd \cite{Lloyd}, a qubit can be represented physically as a simple mode of electromagnetic field trapped in a cavity with high quality factor Q; the total energy is a multiple of $\hslash\omega $. A $CNOT$ logic gate that performs the transformation
\begin{eqnarray}
\left\vert 00\right\rangle _{L} &\mapsto &\left\vert 00\right\rangle _{L}
\notag \\
\left\vert 01\right\rangle _{L} &\mapsto &\left\vert 01\right\rangle _{L}
\notag \\
\left\vert 10\right\rangle _{L} &\mapsto &\left\vert 11\right\rangle _{L}
\notag \\
\left\vert 11\right\rangle _{L} &\mapsto &\left\vert 10\right\rangle _{L},
\end{eqnarray}\label{eq3}
in which the subindex $L$ indicates logic states, can be implemented through the identification
\begin{eqnarray}
\left\vert 00\right\rangle _{L} &\mapsto &\left\vert 0\right\rangle   \notag\\
\left\vert 01\right\rangle _{L} &\mapsto &\left\vert 2\right\rangle   \notag\\
\left\vert 10\right\rangle _{L} &\mapsto &(\left\vert 4\right\rangle +\left\vert 1\right\rangle )/\sqrt{2}  \notag \\
\left\vert 11\right\rangle _{L} &\mapsto &(\left\vert 4\right\rangle
-\left\vert 1\right\rangle )/\sqrt{2},  \label{identif1}
\end{eqnarray}
where the absence of subindex $L$ indicates physical states of the Fock space. In order to show why this occurs, we note that after a time interval $t$, a vector state is given by
\begin{equation}
\left\vert \Psi (t)\right\rangle =e^{-iHt/\hslash }\left\vert \Psi(0)\right\rangle =\underset{n}{\sum }c_{n}e^{-in\omega t}\left\vert
n\right\rangle.
\end{equation}
Considering the evolution of the state until time $t=\pi /\omega $, we have
\begin{eqnarray}
\left\vert \Psi (t=\pi /\omega )\right\rangle  &=&\underset{n}{\sum }c_{n}e^{-in\omega \pi /\omega }\left\vert n\right\rangle =\underset{n}{\sum }c_{n}e^{-in\omega \pi /\omega }\left\vert n\right\rangle   \notag \\
&=&\underset{n}{\sum }c_{n}\cos (-n\pi )\left\vert n\right\rangle =\underset{n}{\sum }c_{n}(-1)^{n}\left\vert n\right\rangle.
\end{eqnarray}
By this time evolution, states $\left\vert 0\right\rangle ,\left\vert2\right\rangle $ and $\left\vert 4\right\rangle $ remain unchanged and $\left\vert 1\right\rangle $ becomes $-\left\vert 1\right\rangle $. Consider then the following superposition

\begin{equation}
\left\vert \Psi \right\rangle _{L}=x^{{\acute{}}}\left\vert 00\right\rangle
_{L}+y^{{\acute{}}}\left\vert 01\right\rangle _{L}+z^{{\acute{}}}\left\vert
10\right\rangle _{L}+w^{{\acute{}}}\left\vert 11\right\rangle _{L},
\end{equation}
with $\left\vert x^{{\acute{}}}\right\vert ^{2}+\left\vert y^{{\acute{}}}\right\vert ^{2}+\left\vert z^{{\acute{}}}\right\vert ^{2}+\left\vert w^{{\acute{}}}\right\vert ^{2}=1$. By using (\ref{identif1}), we have the physical state
\begin{equation}
\left\vert \Psi \right\rangle =x\left\vert 0\right\rangle +y\left\vert1\right\rangle +z\left\vert 2\right\rangle +w\left\vert 4\right\rangle ,
\end{equation}
where
\begin{equation}
\left\{
\begin{array}{l}
x=x^{{\acute{}}} \\
y=\frac{1}{\sqrt{2}}(z^{{\acute{}}}-w^{{\acute{}}}) \\
z=y^{{\acute{}}} \\
w=\frac{1}{\sqrt{2}}(z^{{\acute{}}}+w^{{\acute{}}})
\end{array}
\right.
\end{equation}

According to Thermofield Dynamics and equation (\ref{nova1}), the thermal state\ \ $|\Psi (\beta )\rangle $\ associated with the state $\left\vert \Psi \right\rangle $ is given by
\begin{equation}
\left\vert \Psi (\beta )\right\rangle =x\left\vert 0(\beta )\right\rangle
+y\left\vert 1(\beta )\right\rangle +z\left\vert 2(\beta )\right\rangle
+w\left\vert 4(\beta )\right\rangle ,
\end{equation}
where the thermal states $|n(\beta )\rangle ,n=0,1,2,3,...$ are
\begin{equation}
\left\{
\begin{array}{l}
\left\vert 0(\beta )\right\rangle =\left\vert \beta ;0,\overset{\sim }{0}%
\right\rangle  \\
\left\vert 1(\beta )\right\rangle =a^{\dagger }(\beta )\left\vert 0(\beta
)\right\rangle  \\
\left\vert 2(\beta )\right\rangle =\frac{1}{\sqrt{2}}[a^{\dagger }(\beta
)]^{2}\left\vert 0(\beta )\right\rangle  \\
\left\vert 4(\beta )\right\rangle =\frac{1}{\sqrt{4!}}[a^{\dagger }(\beta
)]^{4}\left\vert 0(\beta )\right\rangle.
\end{array}
\right.
\end{equation}
with $a^{\dagger }(\beta )=U (\beta )a^{\dagger }U(\beta)^{\dagger}$, the thermalized creation operator.

Using the relation \cite{Ademir}
\begin{equation}
a^{\dagger }(\beta )\left\vert 0(\beta )\right\rangle =\frac{1}{u(\beta )}a\left\vert 0(\beta )\right\rangle ,
\end{equation}
where $u(\beta)=\cosh \theta(\beta)$, we can rewrite the above expressions as
\begin{equation}
\left\{
\begin{array}{l}
\left\vert 0(\beta )\right\rangle =\left\vert \beta ;0,\overset{\sim }{0}\right\rangle  \\
\left\vert 1(\beta )\right\rangle =\frac{1}{u(\beta )}a^{\dagger }\left\vert 0(\beta )\right\rangle  \\
\left\vert 2(\beta )\right\rangle =\frac{1}{\sqrt{2}}\frac{1}{u^{2}(\beta )}(a^{\dagger })^{2}\left\vert 0(\beta )\right\rangle  \\
\left\vert 4(\beta )\right\rangle =\frac{1}{\sqrt{4!}}\frac{1}{u^{4}(\beta )}(a^{\dagger })^{4}\left\vert 0(\beta )\right\rangle.
\end{array}
\right.
\end{equation}
Thus, the state $\left\vert \Psi (\beta )\right\rangle $ can be written as
\begin{eqnarray}
\left\vert \Psi (\beta )\right\rangle & = & x\left\vert 0(\beta )\right\rangle +y\frac{1}{u(\beta )}a^{\dagger }\left\vert 0(\beta )\right\rangle +z\frac{1}{\sqrt{2}}\frac{1}{u^{2}(\beta )}(a^{\dagger })^{2}\left\vert 0(\beta)\right\rangle \nonumber \\
&& +w\frac{1}{\sqrt{4!}}\frac{1}{u^{4}(\beta )}(a^{\dagger})^{4}\left\vert 0(\beta )\right\rangle
\end{eqnarray}
The correspondent vector $\langle \Psi (\beta )|$ is
\begin{eqnarray}
\left\langle \Psi (\beta )\right\vert &=&\left\langle 0(\beta )\right\vert x^{\ast }+\left\langle 0(\beta )\right\vert a\frac{1}{u(\beta )}y^{\ast}+\left\langle 0(\beta )\right\vert a^{2}\frac{1}{u^{2}(\beta )}z^{\ast }\frac{1}{\sqrt{2}} \nonumber \\
&& +\left\langle 0(\beta )\right\vert a^{4}\frac{1}{u^{4}(\beta )}w^{\ast }\frac{1}{\sqrt{4!}}
\end{eqnarray}
and the average value of an observable $A$ is given by

\begin{eqnarray*}
&&\left\langle \Psi \left( \beta \right) \right\vert A\left\vert \Psi \left(\beta \right) \right\rangle = \nonumber \\
&=& \left\langle 0(\beta )\right\vert  \{ x^{\ast}+ a\frac{1}{u(\beta )}y^{\ast }+ a^{2}\frac{1}{u^{2}(\beta )}z^{\ast }\frac{1}{\sqrt{2}}+ a^{4}\frac{1}{u^{4}(\beta )}w^{\ast }\frac{1}{\sqrt{4!}}\} \nonumber \\
&& A \{ x + y \frac{1}{u(\beta )}a^{\dagger } + z\frac{1}{\sqrt{2}}\frac{1}{u^{2}(\beta )}(a^{\dagger })^{2} +w\frac{1}{\sqrt{4!}}\frac{1}{u^{4}(\beta )}(a^{\dagger })^{4} \} \left\vert 0(\beta )\right\rangle \nonumber \\
&=&\left\vert x\right\vert ^{2}\left\langle 0\left( \beta \right)
\right\vert A\left\vert 0\left( \beta \right) \right\rangle +\frac{x^{\ast }y%
}{u(\beta )}\left\langle 0\left( \beta \right) \right\vert Aa^{\dagger
}\left\vert 0\left( \beta \right) \right\rangle  \\
&&+\frac{x^{\ast }z}{\sqrt{2}u^{2}(\beta )}\left\langle 0\left( \beta
\right) \right\vert A(a^{\dagger })^{2}\left\vert 0\left( \beta \right)
\right\rangle +\frac{x^{\ast }w}{\sqrt{4!}u^{4}(\beta )}\left\langle 0\left(
\beta \right) \right\vert A(a^{\dagger })^{4}\left\vert 0\left( \beta
\right) \right\rangle  \\
&&+\frac{y^{\ast }x}{u(\beta )}\left\langle 0\left( \beta \right)
\right\vert aA\left\vert 0\left( \beta \right) \right\rangle +\frac{%
\left\vert y\right\vert ^{2}}{u^{2}(\beta )}\left\langle 0\left( \beta
\right) \right\vert aAa^{\dagger }\left\vert 0\left( \beta \right)
\right\rangle  \\
&&+\frac{y^{\ast }z}{\sqrt{2}u^{3}(\beta )}\left\langle 0\left( \beta
\right) \right\vert aA(a^{\dagger })^{2}\left\vert 0\left( \beta \right)
\right\rangle +\frac{y^{\ast }w}{\sqrt{4!}u^{5}(\beta )}\left\langle 0\left(
\beta \right) \right\vert aA(a^{\dagger })^{4}\left\vert 0\left( \beta
\right) \right\rangle  \\
&&+\frac{z^{\ast }x}{\sqrt{2}u^{2}(\beta )}\left\langle 0\left( \beta
\right) \right\vert a^{2}A\left\vert 0\left( \beta \right) \right\rangle +%
\frac{z^{\ast }y}{\sqrt{2}u^{3}(\beta )}\left\langle 0\left( \beta \right)
\right\vert a^{2}Aa^{\dagger }\left\vert 0\left( \beta \right) \right\rangle
\\
&&+\frac{\left\vert z\right\vert ^{2}}{2u^{4}(\beta )}\left\langle 0\left(
\beta \right) \right\vert a^{2}A(a^{\dagger })^{2}\left\vert 0\left( \beta
\right) \right\rangle +\frac{z^{\ast }w}{\sqrt{2}\sqrt{4!}u^{6}(\beta )}%
\left\langle 0\left( \beta \right) \right\vert a^{2}A(a^{\dagger
})^{4}\left\vert 0\left( \beta \right) \right\rangle  \\
&&+\frac{w^{\ast }x}{\sqrt{4!}u^{4}(\beta )}\left\langle 0\left( \beta
\right) \right\vert a^{4}A\left\vert 0\left( \beta \right) \right\rangle +%
\frac{w^{\ast }y}{\sqrt{4!}u^{5}(\beta )}\left\langle 0\left( \beta \right)
\right\vert a^{4}Aa^{\dagger }\left\vert 0\left( \beta \right) \right\rangle
\\
&&+\frac{w^{\ast }z}{\sqrt{2}\sqrt{4!}u^{6}(\beta )}\left\langle 0\left(
\beta \right) \right\vert a^{4}A(a^{\dagger })^{2}\left\vert 0\left( \beta
\right) \right\rangle \nonumber \\
&&+\frac{\left\vert w\right\vert ^{2}}{(\sqrt{4!}%
)^{2}u^{8}(\beta )}\left\langle 0\left( \beta \right) \right\vert
a^{4}A(a^{\dagger })^{4}\left\vert 0\left( \beta \right) \right\rangle .
\end{eqnarray*}%
Using that
\begin{equation}
\left\langle \Psi \left( \beta \right) \right\vert A\left\vert \Psi \left(
\beta \right) \right\rangle =Tr(\rho _{\left\vert \Psi \left( \beta \right)
\right\rangle }A),
\end{equation}%
we have
\begin{eqnarray*}
&&\left\langle \Psi \left( \beta \right) \right\vert A\left\vert \Psi \left(
\beta \right) \right\rangle = \\
&=&\left\vert x\right\vert ^{2}Tr(\rho _{\left\vert 0\left( \beta \right)
\right\rangle }A)+\frac{x^{\ast }y}{u(\beta )}Tr(\rho _{\left\vert 0\left(
\beta \right) \right\rangle }Aa^{\dagger }) \\
&&+\frac{x^{\ast }z}{\sqrt{2}u^{2}(\beta )}Tr(\rho _{\left\vert 0\left(
\beta \right) \right\rangle }A(a^{\dagger })^{2})+\frac{x^{\ast }w}{\sqrt{4!}%
u^{4}(\beta )}Tr(\rho _{\left\vert 0\left( \beta \right) \right\rangle
}A(a^{\dagger })^{4}) \\
&&+\frac{y^{\ast }x}{u(\beta )}Tr(\rho _{\left\vert 0\left( \beta \right)
\right\rangle }aA)+\frac{\left\vert y\right\vert ^{2}}{u^{2}(\beta )}Tr(\rho
_{\left\vert 0\left( \beta \right) \right\rangle }aAa^{\dagger }) \\
&&+\frac{y^{\ast }z}{\sqrt{2}u^{3}(\beta )}Tr(\rho _{\left\vert 0\left(
\beta \right) \right\rangle }aA(a^{\dagger })^{2})+\frac{z^{\ast }w}{\sqrt{4!%
}u^{5}(\beta )}Tr(\rho _{\left\vert 0\left( \beta \right) \right\rangle
}aA(a^{\dagger })^{4}) \\
&&+\frac{z^{\ast }x}{\sqrt{2}u^{2}(\beta )}Tr(\rho _{\left\vert 0\left(
\beta \right) \right\rangle }a^{2}A)+\frac{z^{\ast }y}{\sqrt{2}u^{3}(\beta )}%
Tr(\rho _{\left\vert 0\left( \beta \right) \right\rangle }a^{2}A(a^{\dagger
})^{2}) \\
&&+\frac{\left\vert z\right\vert ^{2}}{2u^{4}(\beta )}Tr(\rho _{\left\vert
0\left( \beta \right) \right\rangle }a^{2}A(a^{\dagger })^{2})+\frac{z^{\ast
}w}{\sqrt{2}\sqrt{4!}u^{6}(\beta )}Tr(\rho _{\left\vert 0\left( \beta
\right) \right\rangle }a^{2}A(a^{\dagger })^{4}) \\
&&+\frac{w^{\ast }x}{\sqrt{4!}u^{4}(\beta )}Tr(\rho _{\left\vert 0\left(
\beta \right) \right\rangle }a^{4}A)+\frac{w^{\ast }y}{\sqrt{4!}u^{5}(\beta )%
}Tr(\rho _{\left\vert 0\left( \beta \right) \right\rangle }a^{4}Aa^{\dagger
}) \\
&&+\frac{w^{\ast }z}{\sqrt{4!}\sqrt{2}u^{6}(\beta )}Tr(\rho _{\left\vert
0\left( \beta \right) \right\rangle }a^{4}A(a^{\dagger })^{2})+\frac{%
\left\vert w\right\vert ^{2}}{(\sqrt{4!})^{2}u^{8}(\beta )}Tr(\rho
_{\left\vert 0\left( \beta \right) \right\rangle }a^{4}A(a^{\dagger })^{4}).
\end{eqnarray*}
Then, with the cyclic property of trace, the density operator becomes%
\begin{eqnarray}
\rho _{\left\vert \Psi \left( \beta \right) \right\rangle } &=&\left\vert
x\right\vert ^{2}\rho _{\left\vert 0\left( \beta \right) \right\rangle }+
\frac{x^{\ast }y}{u(\beta )}a^{\dagger }\rho _{\left\vert 0\left( \beta
\right) \right\rangle }+\frac{x^{\ast }z}{\sqrt{2}u^{2}(\beta )}(a^{\dagger
})^{2}\rho _{\left\vert 0\left( \beta \right) \right\rangle }  \notag \\
&+&\frac{x^{\ast }w}{\sqrt{4!}u^{4}(\beta )}(a^{\dagger })^{4}\rho
_{\left\vert 0\left( \beta \right) \right\rangle }+\frac{y^{\ast }x}{u(\beta
)}\rho _{\left\vert 0\left( \beta \right) \right\rangle }a+\frac{\left\vert
y\right\vert ^{2}}{u^{2}(\beta )}a^{\dagger }\rho _{\left\vert 0\left( \beta
\right) \right\rangle }a  \notag \\
&+&\frac{y^{\ast }z}{\sqrt{2}u^{3}(\beta )}(a^{\dagger })^{2}\rho
_{\left\vert 0\left( \beta \right) \right\rangle }a+\frac{z^{\ast }w}{\sqrt{
4!}u^{5}(\beta )}(a^{\dagger })^{4}\rho _{\left\vert 0\left( \beta \right)
\right\rangle }a \nonumber \\
&+&\frac{z^{\ast }x}{\sqrt{2}u^{2}(\beta )}\rho _{\left\vert 0\left( \beta \right) \right\rangle }a^{2}+\frac{z^{\ast }y}{\sqrt{2}u^{3}(\beta )}a^{\dagger }\rho _{\left\vert0\left( \beta \right) \right\rangle }a^{2}\nonumber \\
&+& \frac{\left\vert z\right\vert ^{2}}{2u^{4}(\beta )}(a^{\dagger })^{2}\rho _{\left\vert 0\left( \beta \right)
\right\rangle }a^{2}+\frac{z^{\ast }w}{\sqrt{2}\sqrt{4!}u^{6}(\beta )}
(a^{\dagger })^{4}\rho _{\left\vert 0\left( \beta \right) \right\rangle
}a^{2}  \notag \\
&+&\frac{w^{\ast }x}{\sqrt{4!}u^{4}(\beta )}\rho _{\left\vert 0\left( \beta \right) \right\rangle }a^{4}+\frac{w^{\ast }y}{\sqrt{4!}u^{5}(\beta )}a^{\dagger }\rho _{\left\vert 0\left( \beta \right) \right\rangle }a^{4}\nonumber \\
&+& \frac{w^{\ast }z}{\sqrt{4!}\sqrt{2}u^{6}(\beta )}(a^{\dagger })^{2}\rho_{\left\vert 0\left( \beta \right) \right\rangle }a^{4}+\frac{\left\vert w\right\vert ^{2}}{(\sqrt{4!})^{2}u^{8}(\beta )}
(a^{\dagger })^{4}\rho _{\left\vert 0\left( \beta \right) \right\rangle}a^{4}.
\end{eqnarray}
Using expression (\ref{rdezerobeta}) we have, explicitly, the mixed state associated with the original thermal state, i.e.:

\begin{eqnarray}
\rho _{\left\vert \Psi \left( \beta \right) \right\rangle } &=&\left\vert
x\right\vert ^{2}k\overset{\infty }{\underset{n=0}{\sum }}
k_{1}^{n}\left\vert n\right\rangle \left\langle n\right\vert +\frac{x^{\ast
}yk}{u(\beta )}\overset{\infty }{\underset{n=0}{\sum }}k_{1}^{n}\sqrt{n+1}
\left\vert n+1\right\rangle \left\langle n\right\vert   \notag \\
&&+\frac{x^{\ast }z}{\sqrt{2}u^{2}(\beta )}k\overset{\infty }{\underset{n=0}{
\sum }}k_{1}^{n}\sqrt{(n+1)(n+2)}\left\vert n+2\right\rangle \left\langle
n\right\vert   \notag \\
&&+\frac{x^{\ast }w}{\sqrt{4!}u^{4}(\beta )}k\overset{\infty }{\underset{n=0}
{\sum }}k_{1}^{n}\sqrt{(n+1)(n+2)(n+3)(n+4)}\left\vert n+4\right\rangle
\left\langle n\right\vert   \notag \\
&&+\frac{y^{\ast }xk}{u(\beta )}\overset{\infty }{\underset{n=0}{\sum }}
k_{1}^{n}\sqrt{(n+1)}\left\vert n\right\rangle \left\langle n+1\right\vert +
\frac{\left\vert y\right\vert ^{2}}{u^{2}(\beta )}k\overset{\infty }{
\underset{n=0}{\sum }}k_{1}^{n}(n+1)\left\vert n+1\right\rangle \left\langle
n+1\right\vert   \notag \\
&&+\frac{y^{\ast }z}{\sqrt{2}u^{3}(\beta )}k\overset{\infty }{\underset{n=0}{
\sum }}k_{1}^{n}(n+1)\sqrt{(n+2)}\left\vert n+2\right\rangle \left\langle
n+1\right\vert   \notag \\
&&+\frac{z^{\ast }w}{\sqrt{4!}u^{5}(\beta )}k\overset{\infty }{\underset{n=0}
{\sum }}k_{1}^{n}(n+1)\sqrt{(n+2)(n+3)(n+4)}\left\vert n+4\right\rangle
\left\langle n+1\right\vert   \notag \\
&&+\frac{z^{\ast }x}{\sqrt{2}u^{2}(\beta )}k\overset{\infty }{\underset{n=0}{
\sum }}k_{1}^{n}\sqrt{(n+1)(n+2)}\left\vert n\right\rangle \left\langle
n+2\right\vert   \notag \\
&&+\frac{z^{\ast }y}{\sqrt{2}u^{3}(\beta )}k\overset{\infty }{\underset{n=0}{
\sum }}k_{1}^{n}(n+1)\sqrt{(n+2)}\left\vert n+1\right\rangle \left\langle
n+2\right\vert   \notag \\
&&+\frac{\left\vert z\right\vert ^{2}}{2u^{4}(\beta )}k\overset{\infty }{
\underset{n=0}{\sum }}k_{1}^{n}(n+1)(n+2)\left\vert n+2\right\rangle
\left\langle n+2\right\vert   \notag \\
&&+\frac{z^{\ast }w}{\sqrt{2}\sqrt{4!}u^{6}(\beta )}k\overset{\infty }{
\underset{n=0}{\sum }}k_{1}^{n}(n+1)(n+2)\sqrt{(n+3)(n+4)}\left\vert
n+4\right\rangle \left\langle n+2\right\vert   \notag \\
&&+\frac{w^{\ast }x}{\sqrt{4!}u^{4}(\beta )}k\overset{\infty }{\underset{n=0}
{\sum }}k_{1}^{n}\sqrt{(n+1)(n+2)(n+3)(n+4)}\left\vert n\right\rangle
\left\langle n+4\right\vert   \notag \\
&&+\frac{w^{\ast }y}{\sqrt{4!}u^{5}(\beta )}k\overset{\infty }{\underset{n=0}
{\sum }}k_{1}^{n}(n+1)\sqrt{(n+2)(n+3)(n+4)}\left\vert n+1\right\rangle
\left\langle n+4\right\vert   \notag \\
&&+\frac{w^{\ast }z}{\sqrt{4!}\sqrt{2}u^{6}(\beta )}k\overset{\infty }{
\underset{n=0}{\sum }}k_{1}^{n}(n+1)(n+2)\sqrt{(n+3)(n+4)}\left\vert
n+2\right\rangle \left\langle n+4\right\vert   \notag \\
&&+\frac{\left\vert w\right\vert ^{2}}{(\sqrt{4!})^{2}u^{8}(\beta )}k\overset
{\infty }{\underset{n=0}{\sum }}k_{1}^{n}(n+1)(n+2)(n+3)(n+4)\left\vert
n+4\right\rangle \left\langle n+4\right\vert ,  \label{matrizde}
\end{eqnarray}
with
\begin{equation}
k=\frac{1}{1+\overset{-}{n}(\beta )}
\end{equation}%
and
\begin{equation}
k_{1}=\frac{\overset{-}{n}(\beta )}{1+\overset{-}{n}(\beta )},
\end{equation}%
where
\begin{equation}
\overset{-}{n}(\beta )=\frac{1}{e^{\beta \omega }-1}.
\end{equation}
Using the density operator $\rho _{\left\vert \Psi \left( \beta \right) \right\rangle }$ we can calculate the fidelity and Wigner function.

\section{Fidelity}

A quantity of interest in Quantum Information\ is the fidelity, which allows, here, to determine the similarity between the original pure state $|\Psi \rangle $ and the thermalized state $|\Psi (\beta )\rangle$.

The fidelity \cite{Uhlmann, Josza} between a pure and a mixed state is given by:
\begin{equation}
F=\sqrt{\left\langle \Psi \right\vert \rho _{\left\vert \Psi \left( \beta
\right) \right\rangle }\left\vert \Psi \right\rangle } .
\end{equation}\label{eq5}
We have, by using (12), that
\begin{eqnarray}
\left\langle \Psi \right\vert \rho _{\left\vert \Psi \left( \beta \right)
\right\rangle }\left\vert \Psi \right\rangle &=&(\left\langle 0\right\vert
x^{\ast }+\left\langle 1\right\vert y^{\ast }+\left\langle 2\right\vert
z^{\ast }+\left\langle 1\right\vert w^{\ast })\rho _{\left\vert \Psi \left(
\beta \right) \right\rangle }(x\left\vert 0\right\rangle +y\left\vert
1\right\rangle +z\left\vert 2\right\rangle +w\left\vert 4\right\rangle )
\notag \nonumber \\
&=&\left\vert x\right\vert ^{2}\left\langle 0\right\vert \rho _{\left\vert
\Psi \left( \beta \right) \right\rangle }\left\vert 0\right\rangle +x^{\ast
}y\left\langle 0\right\vert \rho _{\left\vert \Psi \left( \beta \right)
\right\rangle }\left\vert 1\right\rangle +x^{\ast }z\left\langle
0\right\vert \rho _{\left\vert \Psi \left( \beta \right) \right\rangle
}\left\vert 2\right\rangle \nonumber \\
&& + x^{\ast }w\left\langle 0\right\vert \rho_{\left\vert \Psi \left( \beta \right) \right\rangle }\left\vert
4\right\rangle+y^{\ast }x\left\langle 1\right\vert \rho _{\left\vert \Psi \left( \beta
\right) \right\rangle }\left\vert 0\right\rangle +\left\vert y\right\vert
^{2}\left\langle 1\right\vert \rho _{\left\vert \Psi \left( \beta \right)
\right\rangle }\left\vert 1\right\rangle \nonumber \\
&&+y^{\ast }z\left\langle 1\right\vert \rho _{\left\vert \Psi \left( \beta \right) \right\rangle
}\left\vert 2\right\rangle +y^{\ast }w\left\langle 1\right\vert \rho
_{\left\vert \Psi \left( \beta \right) \right\rangle }\left\vert
4\right\rangle +z^{\ast }x\left\langle 2\right\vert \rho _{\left\vert \Psi \left( \beta
\right) \right\rangle }\left\vert 0\right\rangle \nonumber \\
&&+z^{\ast }y\left\langle 2\right\vert \rho _{\left\vert \Psi \left( \beta \right) \right\rangle
}\left\vert 1\right\rangle +\left\vert z\right\vert ^{2}\left\langle
2\right\vert \rho _{\left\vert \Psi \left( \beta \right) \right\rangle
}\left\vert 2\right\rangle +z^{\ast }w\left\langle 2\right\vert \rho
_{\left\vert \Psi \left( \beta \right) \right\rangle }\left\vert
4\right\rangle   \notag \nonumber \\
&&+w^{\ast }x\left\langle 4\right\vert \rho _{\left\vert \Psi \left( \beta
\right) \right\rangle }\left\vert 0\right\rangle +w^{\ast }y\left\langle
4\right\vert \rho _{\left\vert \Psi \left( \beta \right) \right\rangle
}\left\vert 1\right\rangle +w^{\ast }z\left\langle 4\right\vert \rho
_{\left\vert \Psi \left( \beta \right) \right\rangle }\left\vert
2\right\rangle \nonumber \\
&&+\left\vert w\right\vert ^{2}\left\langle 4\right\vert \rho
_{\left\vert \Psi \left( \beta \right) \right\rangle }\left\vert
4\right\rangle .
\end{eqnarray}\label{eq6}
Then, the fidelity becomes
\begin{eqnarray}
F &=&[|x|^{4}kk_{1}^{n=0}+\frac{|x|^{2}|y|^{2}}{u(\beta )}kk_{1}^{n=0}+\frac{|x|^{2}|z|^{2}}{\sqrt{2}u^{2}(\beta )}kk_{1}^{n=0} \nonumber \\
&&+\frac{|x|^{2}|w|^{2}}{\sqrt{4!}u^{4}(\beta )}kk_{1}^{n=0}+\frac{|x|^{2}|y|^{2}}{u(\beta)}kk_{1}^{n=0}+\frac{|x|^{4}|y|^{2}}{u^{2}(\beta )}kk_{1}^{n=1}+\frac{|y|^{4}}{u^{2}(\beta )}kk_{1}^{n=0} \nonumber \\
&&+\frac{\sqrt{2}\left\vert y\right\vert ^{2}xz}{u(\beta )}kk_{1}^{n=1}+\frac{\left\vert y\right\vert ^{2}\left\vert z\right\vert ^{2}}{u^{3}(\beta )}kk_{1}^{n=0}+\frac{\sqrt{24}\left\vert y\right\vert ^{2}\left\vert w\right\vert ^{2}}{\sqrt{4!}u^{5}(\beta)}kk_{1}^{n=0} \nonumber \\
&&+\frac{\sqrt{2}\left\vert x\right\vert ^{2}\left\vert z\right\vert ^{2}}{u^{2}(\beta )}kk_{1}^{n=0}+\frac{\sqrt{2}x^{\ast }z^{\ast }y^{2}}{u(\beta )}kk_{1}^{n=1}+\frac{\left\vert y\right\vert ^{2}\left\vert z\right\vert ^{2}}{u^{3}(\beta )}kk_{1}^{n=0} \nonumber  \\
&&+\left\vert x\right\vert ^{2}\left\vert z\right\vert ^{2}kk_{1}^{n=2}+\frac{2\left\vert y\right\vert ^{2}\left\vert z\right\vert ^{2}}{u^{2}(\beta)}kk_{1}^{n=1}+\frac{\left\vert z\right\vert ^{4}}{u^{4}(\beta )}kk_{1}^{n=2} \nonumber \\
&&+\frac{\sqrt{6}xw\left\vert z\right\vert ^{2}}{u^{2}(\beta )}kk_{1}^{n=2}+\frac{\sqrt{6}\left\vert z\right\vert ^{2}\left\vert w\right\vert ^{2}}{\sqrt{4!}u^{6}(\beta )}kk_{1}^{n=0}+\frac{\sqrt{24}xw^{\ast }}{\sqrt{4!}u^{4}(\beta )}kk_{1}^{n=0} \nonumber \\
&&+\frac{\sqrt{24}\left\vert z\right\vert ^{2}\left\vert w\right\vert ^{2}}{\sqrt{4!}u^{5}(\beta )}kk_{1}^{n=0}+\frac{\sqrt{6}x^{2}z^{2}w^{\ast }}{u^{4}(\beta )}kk_{1}^{n=2}+\frac{2\sqrt{6}\left\vert z\right\vert^{2}\left\vert w\right\vert ^{2}}{\sqrt{4!}u^{4}(\beta )}kk_{1}^{n=0} \nonumber  \\
&&+\left\vert x\right\vert ^{2}\left\vert w\right\vert ^{2}kk_{1}^{n=4}+\frac{4\left\vert y\right\vert ^{2}\left\vert w\right\vert ^{2}}{u^{2}(\beta)}kk_{1}^{n=3}+\frac{\left\vert z\right\vert ^{2}\left\vert w\right\vert ^{2}}{2u^{4}(\beta )}kk_{1}^{n=2}+\frac{24\left\vert w\right\vert ^{4}}{4!u^{8}(\beta )}kk_{1}^{n=0}, \nonumber \\
\end{eqnarray}
where we have used eq.(22). For graphical analysis, we plot the fidelity $F$ in terms of $\overset{-}{n}(\beta)$. In order to analyze the $F$ for a particular state, $|\Psi(\beta) \rangle$, we consider in eq.(28) $x=0.2,y=0.3,z=0.6$, $w=\sqrt{0.51}$ and the result given by Khanna \textit{et al} \cite{Ademir}
\begin{equation}
u(\beta )=\sqrt{1+\overset{-}{n}(\beta )}.
\end{equation}

\begin{figure}[h]
\caption[Fidelity]{Fidelity (with respect to the pure state $| \Psi \rangle$) as a function of $\overset{-}n \equiv \overset{-}{n}(\beta).$}
\label{Fig:FidelidadeOH}\centering
\includegraphics[width=10cm,height=7cm]{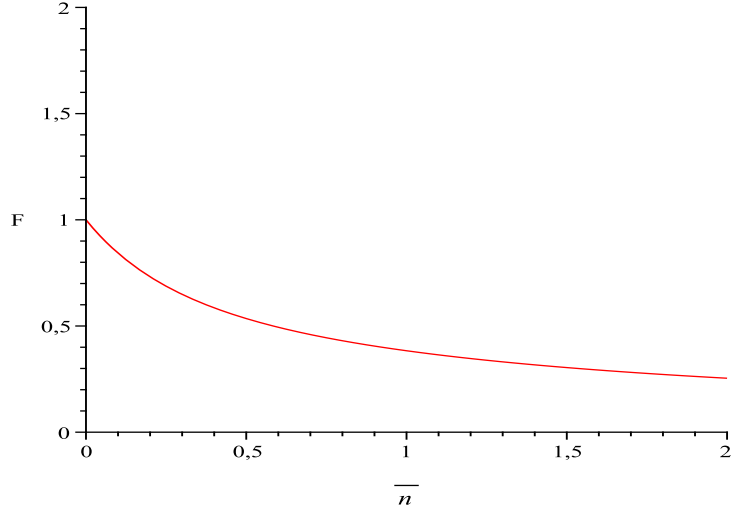}
\end{figure}

Figure \ref{Fig:FidelidadeOH} gives the behavior of $F$. We note that when $T\rightarrow 0$ $(\overset{-}{n}(\beta)\rightarrow 0)$, the fidelity is close to value $F=1$ and its value decreases for $T>0$, that is, the temperature causes the system to move away from its original state so as to be characterized by a statistical mixture of pure states.

\section{ Mandel Parameter}

Subpoissonian statistics of photons is a characteristic of electromagnetic field state  which has not a classical analogue. One way to investigate its occurrence is through the Mandel parameter \cite{Mandel}. The Mandel parameter is defined in terms of a thermalized state by
\begin{equation}
Q(\beta )=\frac{\left\langle \Psi (\beta )\right\vert \overset{\wedge }{N^{2}
}\left\vert \Psi (\beta )\right\rangle -\left\langle \Psi (\beta
)\right\vert \overset{\wedge }{N}\left\vert \Psi (\beta )\right\rangle
^{2}-\left\langle \Psi (\beta )\right\vert \overset{\wedge }{N}\left\vert
\Psi (\beta )\right\rangle }{\left\langle \Psi (\beta )\right\vert \overset{
\wedge }{N}\left\vert \Psi (\beta )\right\rangle },
\end{equation}
with the number operator
\begin{eqnarray}
\overset{\wedge }{N} &=&a^{\dagger }a \nonumber \\
&=&u^{2}(\beta )a^{\dagger }(\beta )a(\beta )+u(\beta )v(\beta )a^{\dagger
}(\beta )\tilde{a}^{\dagger }(\beta )  \notag \\
&&+u(\beta )v(\beta )\tilde{a}(\beta )a(\beta )+v^{2}(\beta )\tilde{a}(\beta
)\tilde{a}^{\dagger }(\beta ).
\end{eqnarray}
where we have used the results obtained by Khanna \textit{et al} \cite{Ademir}
\begin{equation*}
a=u(\beta)a(\beta )+v(\beta )\tilde{a}^{\dagger }(\beta )
\end{equation*}
and
\begin{equation*}
a^{\dagger}=u(\beta )a^{\dagger }(\beta )+v(\beta )\tilde{a}(\beta )
\end{equation*}

\begin{figure}[h]
\caption[Mandel parameter]{Mandel parameter, considering $x=0.2, y=0.3, z=0.6$ and $w=\sqrt{0.51}$, as a function of $\overset{-}n \equiv \overset{-}n(\beta)$.}
\label{Fig:MandelOH}\centering
\includegraphics[width=10cm,height=7cm]{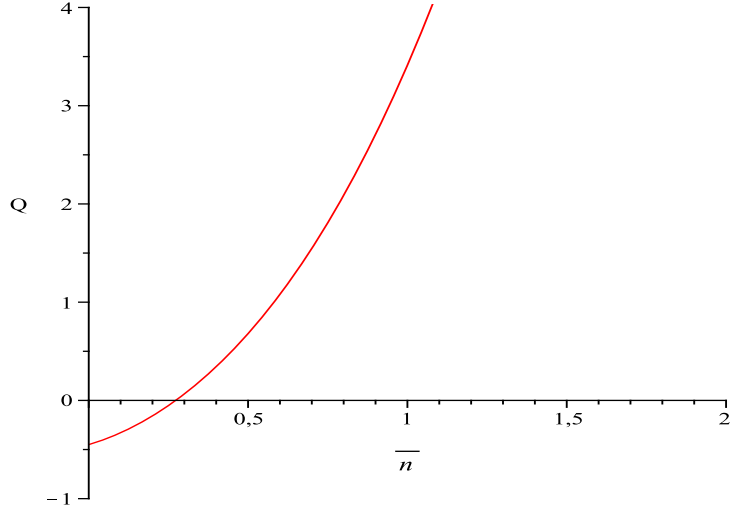}
\end{figure}

Using eq.(18) and eq.(31), we have
\begin{eqnarray}
\left\langle \Psi (\beta )\right\vert \overset{\wedge }{N}\left\vert \Psi
(\beta )\right\rangle  &=&\left\vert x\right\vert ^{2}v^{2}(\beta
)+\left\vert y\right\vert ^{2}[u(^{2}\beta )+v^{2}(\beta )]  \notag \\
&&+\left\vert z\right\vert ^{2}[2u^{2}(\beta )+v^{2}(\beta )]+\left\vert
w\right\vert ^{2}[4u^{2}(\beta )+v^{2}(\beta )]
\end{eqnarray}
and
\begin{eqnarray*}
\left\langle \Psi (\beta )\right\vert \overset{\wedge }{N^{2}}\left\vert
\Psi (\beta )\right\rangle  &=&(\left\vert x\right\vert ^{2}+\left\vert
y\right\vert ^{2}+4\left\vert y\right\vert ^{2}+7\left\vert z\right\vert
^{2}+8\left\vert w\right\vert ^{2})u^{2}(\beta )v^{2}(\beta ) \\
&&+(\left\vert x\right\vert ^{2}+\left\vert w\right\vert ^{2}+\left\vert
y\right\vert ^{2}+\left\vert z\right\vert ^{2})v^{4}(\beta ) \\
&&+(\left\vert y\right\vert ^{2}+4\left\vert z\right\vert ^{2}+16\left\vert w\right\vert
^{2})u^{4}(\beta ).
\end{eqnarray*}%
Then the Mandel parameter is given by
\begin{eqnarray}
Q &=&[(c_{6}-c_{4})u^{2}(\beta )v^{2}(\beta )+(c_{7}-c_{3})v^{4}(\beta
)+(c_{8}-c_{5})u^{4}(\beta )  \notag \\
&&-c_{1}v^{2}(\beta )-c_{2}u^{2}(\beta )]/(c_{1}v^{2}(\beta
)+c_{2}u^{2}(\beta )],
\end{eqnarray}%
with coefficients $c_{1},c_{2},c_{3},c_{4},c_{5},c_{6},c_{7},c_{8},c_{9}$ given by
\begin{eqnarray}
c_{1} &=&\left\vert x\right\vert ^{2}+\left\vert y\right\vert
^{2}+\left\vert z\right\vert ^{2}+\left\vert w\right\vert ^{2}  \notag \\
c_{2} &=&\left\vert y\right\vert ^{2}+2\left\vert z\right\vert
^{2}+4\left\vert w\right\vert ^{2}  \notag \\
c_{3} &=&\left\vert x\right\vert ^{4}+2\left\vert x\right\vert
^{2}\left\vert y\right\vert ^{2}+2\left\vert x\right\vert ^{2}\left\vert
z\right\vert ^{2}+2\left\vert x\right\vert ^{2}\left\vert w\right\vert
^{2}+\left\vert y\right\vert ^{4}+2\left\vert y\right\vert ^{2}\left\vert
z\right\vert ^{2}  \notag \\
&&+2\left\vert y\right\vert ^{2}\left\vert w\right\vert ^{2}+\left\vert
z\right\vert ^{4}+2\left\vert z\right\vert ^{2}\left\vert w\right\vert
^{2}+\left\vert w\right\vert ^{4}  \notag \\
c_{4} &=&\left\vert x\right\vert ^{2}\left\vert y\right\vert ^{2}+\left\vert
y\right\vert ^{4}+3\left\vert y\right\vert ^{2}\left\vert z\right\vert
^{2}+5\left\vert y\right\vert ^{2}\left\vert w\right\vert ^{2}+2\left\vert
x\right\vert ^{2}\left\vert z\right\vert ^{2}+2\left\vert z\right\vert ^{4}
\notag \\
&&+6\left\vert z\right\vert ^{2}\left\vert w\right\vert ^{2}+4\left\vert
x\right\vert ^{2}\left\vert w\right\vert ^{2}+4\left\vert w\right\vert ^{4}
\notag \\
c_{5} &=&\left\vert y\right\vert ^{4}+4\left\vert y\right\vert
^{2}\left\vert z\right\vert ^{2}+8\left\vert y\right\vert ^{2}\left\vert
w\right\vert ^{2}+4\left\vert z\right\vert ^{4}+16\left\vert z\right\vert
^{2}\left\vert w\right\vert ^{2}+16\left\vert w\right\vert ^{4}  \notag \\
c_{6} &=&\left\vert x\right\vert ^{2}+\left\vert y\right\vert
^{2}+4\left\vert y\right\vert ^{2}+7\left\vert z\right\vert ^{2}+8\left\vert
w\right\vert ^{2}  \notag \\
c_{7} &=&\left\vert x\right\vert ^{2}+\left\vert y\right\vert
^{2}+\left\vert w\right\vert ^{2}+\left\vert z\right\vert ^{2}  \notag \\
c_{8} &=&\left\vert y\right\vert ^{2}+4\left\vert z\right\vert
^{2}+16\left\vert w\right\vert ^{2}.
\end{eqnarray}
Considering the parameters $x=0.2,y=0.3,z=0.6$ and $w=\sqrt{0.51},$the behavior of the Mandel parameter as a function of $\overset{-}{n}(\beta )$ can be seen in Figure \ref{Fig:MandelOH}. We note that for $\overset{-}{n}(\beta )<0.3,\overset{-}{n}(\beta)=0.3,\overset{-}{n}(\beta)>0.3$  we have subpoissonian, poissonian and superpoissonian statistics, respectively. Hence, as \ $\overset{-}{n}(\beta)$\ depends on the temperature, we can say that increasing temperature decreases the degree of non-classicality of the system.

\section{Wigner Function}

\begin{figure}[tbh]
    \caption[Wigner function]{Wigner function for $\overset{-}n(\beta)=0.1$ considering the parameters $x=0.2, y=0.3, z=0.6$ and $w=\sqrt{0.51}$.}
    \label{Fig:Wigner01OH}\centering
    \includegraphics[width=10cm,height=7cm]{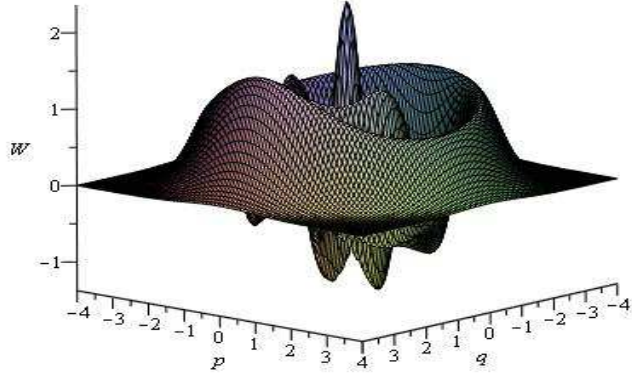}
\end{figure}

\begin{figure}[tbh]
    \caption[Wigner function]{Wigner function for $\overset{-}n(\beta)=10$ considering the parameters $x=0.2, y=0.3, z=0.6$ and $w=\sqrt{0.51}$.}
    \label{Fig:Wigner10OH}\centering
    \includegraphics[width=10cm,height=7cm]{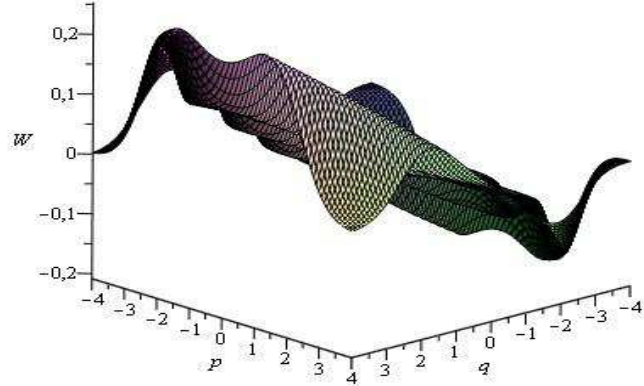}
\end{figure}

Non-classical properties of a physical system can be studied by using the Wigner (quasi-probability distribution) function. An advantage of this approach is that it associates a quantum state in Hilbert space with a real-function $W(q,p)$ where (q, p) are coordinates of a phase space manifold \cite{Kim}. Indeed the Wigner function is introduced by using a kind of Fourier transform of the density matrix and in the present case, we have
\begin{equation}
W(p,q;\beta )=\int_{-\infty }^{\infty }dv\exp (\frac{i}{\hbar}pv)\left\langle q-\frac{v}{2}\right\vert \rho _{\left\vert \Psi (\beta
)\right\rangle }\left\vert q+\frac{v}{2}\right\rangle.
\end{equation}
where $\rho _{\left\vert \Psi (\beta )\right\rangle }$ is given by expression (\ref{matrizde}). In consequence, we have in eq.(35) the following kind of terms:

\begin{equation}
(a) \phantom o \sum_i \sum_n^{\infty} T_i(n)| n + Y_i\rangle \langle n+Y_i |,
\end{equation}
where $Y_i$ can take the values 1, 2 and 4;

\begin{equation}
(b) \phantom o \sum_n^{\infty} T(n)| n \rangle \langle n + Y |,
\end{equation}
where $Y$ can take the values 2 and 4;

\begin{equation}
(c) \phantom o \sum_n^{\infty} T(n)| n +Y \rangle \langle n |,
\end{equation}
where $Y$ can take the values 1 and 2, and

\begin{equation}
(d) \phantom o \sum_n^{\infty} T(n)| n+Y \rangle \langle n + Z |,
\end{equation}
where $(Y,Z)$ can take the values (2,4), (1,4), (4,2), (1,2), (4,1) and (2,1). In relations (36)-(39) $T_i(n)$ and $T(n)$ are numerical coefficients. After some algebraic development we obtain for complete Wigner function:

\begin{eqnarray}
W(p,q;\beta ) &=&k\exp \left[ -\left( \frac{q^{2}}{b^{2}}+\frac{b^{2}p^{2}}{\hbar ^{2}}\right) \right] \underset{n=0}{\overset{\infty }{\sum }}K_{1}^{n}(-1)^{n}\left\{2x^{2}L_{n}\left[ 2\left( \frac{q^{2}}{b^{2}}+\frac{b^{2}p^{2}}{\hbar ^{2}}\right) \right] \right.  \notag \\
&&-\frac{2y^{2}}{u(\beta )}(n+1)L_{n+1}\left[ 2\left(\frac{q^{2}}{b^{2}}+\frac{b^{2}p^{2}}{\hbar ^{2}}\right)\right]   \notag \\
&&+\frac{z^{2}(n+1)(n+2)}{u^{4}(\beta )}L_{n+2}\left[ 2\left(\frac{q^{2}}{b^{2}}+\frac{b^{2}p^{2}}{\hbar ^{2}}\right)\right]   \notag \\
&&+\frac{2w^{2}(n+1)(n+2)(n+3)(n+4)}{(\sqrt{4!})^{2}u^{8}(\beta )}L_{n+4}\left[2\left(\frac{q^{2}}{b^{2}}+\frac{b^{2}p^{2}}{\hbar ^{2}}\right)\right]   \notag
\\
&&+\frac{4\sqrt{2}xy}{u(\beta )}\frac{q}{b}L_{n}^{1}\left[ 2\left( \frac{q^{2}}{b^{2}}+\frac{b^{2}p^{2}}{\hbar ^{2}}\right) \right]   \notag \\
&&+\frac{4\sqrt{2}xz}{u^{2}(\beta )}\left( \frac{q^{2}}{b^{2}}-\frac{b^{2}p^{2}}{\hbar ^{2}}\right) L_{n}^{2}\left[ 2\left( \frac{q^{2}}{b^{2}}+\frac{b^{2}p^{2}}{\hbar ^{2}}\right) \right]   \notag \\
&&+\frac{4\sqrt{6}xw}{3u^{4}(\beta )}\left( \frac{q^{2}}{b^{2}}+\frac{b^{2}p^{2}}{\hbar ^{2}}-\frac{6q^{2}p^{2}}{\hbar ^{2}}\right) L_{n}^{4}\left[2\left( \frac{q^{2}}{b^{2}}+\frac{b^{2}p^{2}}{\hbar ^{2}}\right) \right]
\notag \\
&&-\frac{4(n+1)yz}{u^{3}(\beta )}\frac{q}{b}L_{n+1}^{1}\left[ 2\left(\frac{q^{2}}{b^{2}}+\frac{b^{2}p^{2}}{\hbar ^{2}}\right)\right]   \notag \\
&&+\frac{4\sqrt{3}(n+1)yw}{3u^{5}(\beta )}\left( \frac{q^{3}}{b^{3}}-\frac{3qp^{2}b}{\hbar ^{2}}\right) L_{n+1}^{3}\left[ 2\left(\frac{q^{2}}{b^{2}}+\frac{b^{2}p^{2}}{\hbar ^{2}}\right)\right]   \notag \\
&&+\frac{2\sqrt{3}(n+1)(n+2)wz}{3u^{6}(\beta)}\left( \frac{q^{2}}{b^{2}}-\frac{p^{2}b^{2}}{\hbar ^{2}}\right)   \notag \\
&&\left. \times L_{n+2}^{2}\left[ 2\left(\frac{q^{2}}{b^{2}}+\frac{b^{2}p^{2}}{\hbar ^{2}}\right)\right]\right\}.  \label{fwcompleta}
\end{eqnarray}
where $L_{m}^{n-m},n\geq m$ are associated Laguerre function \cite{Arfken}. Wigner functions  for $\overset{-}{n}(\beta)=0.1$ and $\overset{-}{n}(\beta)=10$, by considering the parameters  $x=0.2,y=0.3,z=0.6$ and $w=\sqrt{0.51}$, are illustrated in Figure \ref{Fig:Wigner01OH} and Figure \ref{Fig:Wigner10OH}, respectively. We note that if the temperature increases (note that the temperature for $\overset{-}{n}(\beta)=10$ is greater than for $\overset{-}{n}(\beta)=0.1$) negative values for the Wigner function become rare, and this indicates that  non-classical properties of the system are lost.

\newpage
\section{Conclusions and perspectives}

The effect of the environment (reservoir or bath) on the evolution of a physical system is a subject of actual interest and has  received contributions of several authors \cite{louisell, caldeira, zurek1, zurek2}.  In fact, no classical or quantum system is perfectly isolated and hence it is important to analyze consequences of the  bath or reservoir-system interaction. In particular in the case of quantum computing  the temperature can modify non-classical properties  and affect the state of qubit corrupting the desired evolution of the system. In this communication, we show that the action of the thermalized quantum logic gate  on the thermalized state is equivalent to thermalization of the state that arise from the application of the non-thermalized quantum logic gate and we have investigated the effect of temperature on a mixed state associated to a system capable of implementing a \textit{CNOT} quantum logic gate. A way of implementing such a  logic gate is by using a representation of the qubit states as elements of the Fock space of a bosonic system. In our  analysis we have considered the Thermofield Dynamics. Here, the temperature acts as a quantum noise on pure states, making them a statistical mixture. By calculating the fidelity between the pure state  $\left\vert \Psi\right\rangle $ and the mixed state associated to thermalized state $\left\vert \Psi (\beta )\right\rangle $, we have obtained a measure of distance between these states as a function of temperature. The degree of non-classically for these states was also investigated through Mandel parameter allowing the determination of temperature ranges for which the statistics of the bosons becomes subpoissonian, poissonian and superpoissonian. We have obtained that for $0<\overset{-}{n}(\beta )<0.3$, it is found subpoissonian statistics and high fidelity, $F > 0.7$. In consequence the $CNOT$ gate can probably be done more efficiently for these values of $\overset{-}{n}(\beta ).$ Finally, we have calculated the Wigner function, allowing an analysis of the thermal state in phase space, indicating the decrease of non-classical properties of the system with increasing temperature. A similar analysis can be realized to other qubit states of interest in quantum computation and quantum information. Works along this line are in progress.

\end{document}